# On the physical and circuit-theoretic significance of the Memristor

*Emanuel Gluskin*

**Kinneret College in the Jordan Valley (on the Sea of Galilee), Tzemach, 15132 Israel**

**http://www.ee.bgu.ac.il/~gluskin/, gluskin@ee.bgu.ac.il**

*Abstract*—**It is observed that the inductive and capacitive features of the memristor reflect (and are a quintessence of) such features of *any* resistor. The very presence of the voltage and current state variables,** *associated by their electrodynamics sense with electrical and magnetic field*s**, in the resistive characteristic $v = f(i)$, forces any resister to accumulate some magnetic and electrostatic fields and energies around itself, i.e. *L* and *C* elements are always present. From the *circuit-theoretic* point of view, the role of the memristor is seen, first of all, in the elimination of the use of a unique $v(i)$. This makes circuits with hysteresis characteristics relevant, and also suggests that the concept of memristor should influence the basic problem of definition of nonlinearity. Since the memristor mainly originates from the resistor, it was found necessary to overview some unusual cases of resistive circuits. The present opinion is that the framework of basic circuit theory and its connection with applications should be logically expanded in order to naturally include the new element.**

## I.  INTRODUCTION

*1.1 General.*  Predicted by L.O. Chua in [1] (see also [2,3]), and already technologically realized [4], the *memristor* (i.e. "resistor with memory") and memristive circuits have became a popular research topic, e.g. [5-18] and references therein.  The original purpose [1] of this specific circuit element is to connect an electrical charge with a magnetic flux, which is outside the borders of the use of the resistive characteristic given only by an equality $v = f(i)$, with a unique $f(.)$.  The element must be nonlinear (NL), remembering (i.e. influenced by) the integral charge passed via it.  The connection of a nonlinearity with a memory is known in magnetic [19], ferroelectric [20,21] and discharge lamps' [22] characteristics, where the hysteresis curves, necessarily associated both with memory and nonlinearity, arise, and now we have a new basic element to be thoroughly discussed from the physical and circuit/system theory points of view.  As is pedagogically stressed in [23], every scientific discovery not only opens some new ways for the future, but also allows us to better understand what was done in the past, and the discovery of the memristor requires the common (usual) resistor to be reconsidered.  Though [1] and to a degree [2] give a formal definition of the memristor, the framework of the basic circuit theory and its connection with applications should be enlarged in order to include the new element naturally.

*1.2 A comment on the application of* **[4].**  Many years have passed after [1] until [4] aroused real interest in memristors, and thus [4] deserves a special attention.  The application of the memristor found in [4] is a two-state resistive switch {$R_{YES}$, $R_{NO}$} such that $R_{NO} \sim 10^4 R_{YES}$, which is a good switching.  The tendency in [4] of using the memristor-switch for improving the speed of the computer operation is a specific and attractive direction that immediately led to the appearance of many other research works.  It is worth considering that the target of [4] requires the values of $R_{YES}$ and $R_{NO}$ to exist very reliably.  This presents some technological challenges, already reflected in [4-18].  (Recall, to the point, that after being forgotten for many years, Boolean {YES, NO} logic received wide applications *only after* the appearance of computers, in which, in the electronics realization, the "NO"-element are as well physically defined and as reliable, as the "YES-"element.)  Noticeably, any hysteresis-type voltage-current (*v-i*) characteristics with very different slopes of d$v$/d$i$, including the curves appearing in Section III, have to be considered in view of the requirement of reliability, which means that the shapes of the hysteresis curves-cycles have to be very stable in the whole range of the frequency operation.  This stability is usually obtained by a clear specification of some of the system's quantum properties, i.e. of the precise discrete values of some physical variables of the system.  However, the fact that any such element is not a good voltage source (that would provide the needed stability of the realization of the values of d$v$/d$i$ and make their use be technically simple), is a problem for applications.

*1.3 The role of the electromagnetic fields*. According to our target of over-viewing some theoretical points that should not be missed in the field, Section II touches, after [24], the fact that the electrical and magnetic fields, which are organic features of *C* and *L* elements, are accumulated in the space around *any conductive element*.  These physical fields are *necessary* for the power losses in the element to occur (even in the dc process), and the accumulated energy associated with the fields represents some memory.  This physical argument cannot be missed, because as a rule, the hysteresis-type *v-i* characteristics, often used in the consideration of memristors [5-18,25], possess some *L* or *C* properties.  The focus on the dc state of a lumped circuit, allows one to better compare the theoretical roles of the memristor and of the common resistor.  That this topic is relevant to the well-known frequency limits of applications of Kirchhoff's equations [26,27] does not mean that this topic is redundant here; it just further stresses the need of the analysis of the fields, and of the electromagnetic interferences, because the suggestion of [4] to increase the speed of the operation of computers means increase of the frequency range.  All this is connected not only with the reliability of the logical operations; we make here some heuristic points for basic circuit theory.

## II. THE COMMON RESISTOR AND THE ENERGY FLOW

Circuit theory [26,27] defines the resistor as a one-port described by the voltage-current one-to-one relation

$$v = f(i), \qquad (1)$$

though this relation need not be given explicitly, that is (1) can be more generally given as some curve $f(v,i) = 0$, in the *v*-i plane. Practical engineers often tend to understand the term "resistor" only as related to the linear dependence $v = Ri$ where the constant coefficient $R$ is the *resistance.*

An immediate observation (unfortunately not appearing in textbooks) is that since (1) is formulated in terms of the capacitor's and inductor's *state variables* (*v* for *C*, and *i* for *L*), *no working resistor can exist without some associated capacitance and inductance, which are some energy-accumulating elements having some memory.*

Indeed, for any flowing current *i* (may be a *direct current*) there is [28] a magnetic field ***H*** around the wire (the element), and thus there is the magnetic energy $\sim H^2$ associated with some inductance. Similarly, if we have a voltage drop *v* on an element, then we have an electrical field ***E*** of which *v* is an integral measure, and since the tangential ***E*** is continuous on the surface of the conductor [28,29], it exists also outside the conductor, and electrostatic energy $\sim E^2$, as that of a capacitor, is accumulated around the conductor. Of course, the fields and the energies exist also inside the conductor.

Furthermore, since the current flows in the direction of ***E***, ***E*** appears to be perpendicular to ***H*** around the conductor, so that the nonzero Pointing vector (the vector product) ***S*** = [***E***,***H***] is directed towards the conductor, and the energy comes from the outer space to the element; in this way it is dissipated in the conductor (i.e. is heating it) as instantaneous power $p = vi$ (for the linear case, $Ri^2$) supplied to the conductor.

Remark 1: For a cylindrical form of the conductor, the direct proof is very simple. The electrical power flowing into the conductor is $sS$ where $s$ is the relevant part of the conductor's surface. Using also the length *l* of the cylinder and its radius *r*, we have for the flow of the energy into the conductor

$$sS = sEH = (l2\pi r)(\frac{v}{l})(\frac{i}{2\pi r}) = vi \qquad (2)$$

i.e. precisely the consumed power *p*. The dependence of *v* on *i* is not important here, the conductor can be electrically nonlinear in any degree.

Thus, even the state with the established *dc* current, is, essentially, a *field problem* [38] in which the surface of the resistor plays the role of the boundary of the space with the field. Figure 1 (used in [24]) illustrates the argument.

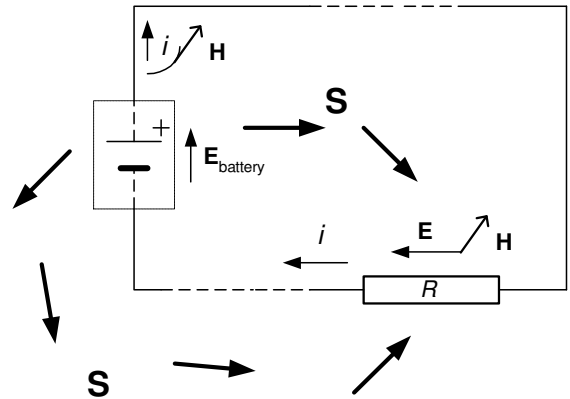

Fig. 1: The energy (power) enters the resistor from outside because of the fields **E** and **H**, and thus $p = p_R = Ri^2$ (or, more generally, $vi$) is obtained. The simplest dc state is in the focus. See also the references in [24], in which the field analysis is presented in detail and for any frequency. One can see here, in particular, a Thevenin equivalent of a more complicated circuit.

*Thus, by the very sense of the variables v and i defining the common resistor, there always is some accumulated electromagnetic energy. In other words, there are a capacitance and an inductance organically associated with any resistor, in any circuit.*

Of course, one should *not* interpret these absolutely necessary capacitance and inductance as parasitic elements, because no parasitic element defines the *basic properties* of a circuit (here, the value of *p*); any resistor has to consume the power as is described.

From the positions of field theory, the existence of both electrical and magnetic fields *in the dc state* is associated with the fact that the boundary is conductive, i.e. there cannot be *v* without *i*, i.e. ***E*** without ***H***. It would be interesting to consider the mutual influence of the circuit and field problems in more detail; for instance, the cases of superconductive elements (when only some internal resistance of the battery is present, i.e. the battery becomes a current-source), several parallel conductors, etc..

**2.1. The argument of the theory of relativity**. Regarding Fig. 1, it is also important to note that if we apply Ohm's law to a series connection of so many resistors that the length of the connection is many kilometers, -- then, applying the battery to the chain and assuming that the *momental (instantaneous)* relation $v = Ri$ (or any other $v = f(i)$) is correct, we can instantaneously transfer the *signal as the current* (i.e. the information about connecting the source) over the whole distance, *which obviously contradicts the theory of relativity stating that no signal can propagate quicker than electromagnetic waves.*

It thus becomes clear that after closing the switch there is an initial electromagnetic wave, propagating from the battery and leading to a finally established process with the dc current, while at this very dc state *there is a steady flow of energy from the battery to the resistor via the outer space*, associated with an inductance and a capacitance. $\vec{S} \neq \vec{0}$!

Thus, the fields and the *L* and *C* properties of the common resistor described by (1) are required by the theory of relativity.

The connection with memristors is seen if one notices that any non-unique function (and we indeed avoid any unique $v = f(i)$!) implies some memory associated with the fields, the energy, and the time delays. Indeed, it is impossible to choose one of the two values of a function, given at the same value of the argument, not knowing which one is obtained first, which is the memory and the development of a process in time.

### III. ON THE MATHEMATICAL (CIRCUIT DESCRIPTION) SIDE

Another immediate observation regarding (1) is that in order to describe hysteresis elements, i.e. to move from the unique characteristic (1) to a non-unique one, it may be sufficient to include $di/dt$ in (1).  The fact that the use of $di/dt$ connects the non-uniqueness with the memory, can be clearly seen from the system

$$v = f(i, \lambda)$$
$$\frac{di}{dt} = \lambda$$

in which we consider $\lambda$ as a parameter ranging along some family $v = f(i)$.  Simultaneously, there is some necessarily accumulated energy, associated with $di/dt$ (i.e. with some voltage, see below) which is an integral measure expressing the same memory.

The relevant function of the two variables $v = f(i, \frac{di}{dt})$ may have the very simple form of the sum of the two functions of single variables, $v = f(i) + g(\frac{di}{dt})$, but we require $f(.)$ to be *singular*, i.e. to have jumps by itself, or in some of its derivatives, which causes Fourier expansion/series of the time-function $\psi(t) \equiv f(i(t))$ to be infinite.  Some such models connecting $v(t)$ with $i(t)$ and $di/dt$ are considered in detail below, and the singularity of $f(.)$ appears to be very important.

The singularity of a model-characteristic is not encouraged, although not prohibited, by the following (introduced in [2] and then used, e.g., in [15]), description of memristive 1-ports:

$$v = R(\vec{x}, i)i$$
$$\frac{d\vec{x}}{dt} = \vec{f}(\vec{x}, i) \qquad (3)$$

where $\vec{x}$ is the vector of the internal state variables of the memristor.

The memory is reflected by (3) because of the time-derivative; and solutions of differential equations depend on the initial conditions, i.e. on the past.  The scalar function $R$ (the "memristance") in (3) generally depends on several variables as well as on the history of the process.

The presence of the variables $v$ and $i$ in (3), as well as in the model $v = f(i) + g(\frac{di}{dt})$, makes the physical comment of Section II also relevant to the memristors. From the mathematical point of view, it is important that in order to move from a common resistor to a memristor, *one* variable included in *R* is sufficient.   Reference [30] considers reduction (via some functions' superposition) of the number of arguments from a function of many variables to functions of a smaller number of variables.  Since it is possible to derive one higher order equation for one variable from a system of first order equations for several variables (the second line of (3)), it is interesting to know when *R*(.,*i*) can include only *i*.

***3.1 Some additional comments on*** **(3)**.  In order to clarify how equations (3) reflects the basic definition [1] of the new element which connects magnetic flux

$$\psi(t) = \int_t v(t)dt$$

with electrical charge

$$q(t) = \int_t i(t)dt,$$

we observe, with some simplification, that from the second equation of (3), $\vec{x}(t)$ should depend on $\int_t i(t)dt = q(t)$. Then, from the first equation,

$$v(t) = F(q(t),i)i$$

with some function *F*(.,.), and

$$\psi(t) = \int_t v(t)\,dt = \int_t F(q(t),i)\,i\,dt = \int_q F(q,i)\,dq.$$

Thus, if $i(t) = dq/dt$, included in *F*, can be directly expressed via $q$ (that is, some differential equation for $q(t)$ can be derived), then we have a direct connection between $\psi$ and $q$. Thus, according to model (3), the *charge variable* has to be well defined.

We can, however, move to the dual situation of:

$$i = G(\vec{x},v)v \qquad (3a)$$

$$\frac{d\vec{x}}{dt} = \vec{f}(\vec{x},v) \qquad (3b)$$

Then, having $\vec{x}(t)$ dependent on $\int_t v(t)dt = \psi(t)$, we know from the first equation that $i(t) = F(\psi(t),v)v$, and

$$q(t) = \int_t i(t)\,dt = \int_t F(\psi,v)\,v\,dt = \int_\psi F(\psi,\frac{d\psi}{dt})\,d\psi.$$

In this case, in order to connect between $q$ and $\psi$, a good description of the flux variable (i.e. a differential equation for $\psi(t)$, expressing $d\psi/dt$ via $\psi$) is needed.

According to the original idea of [1], in the actual realization of the memristor in [4] the resistance depends on the charge that has passed via the same element. However, recalling that from the matrix point of view, the well-known *dependent sources* [26,27] are just some "non-diagonal resistors", one notes that the element's resistance can be dependent on the charge that has passed via *another* element of the same 1-port. Indeed, in (3), the initial conditions, associated with the memory, can relate to some *x* that need not be *v* or *I*; thus a "dependent memristor" can be theoretically obtained. However, there is a problem, since dependent sources have internal sources of energy, while the memristor is basically defined as a passive element. This means that model (3) can be problematic in the energetic sense, which shows that the mathematical description of the memristive systems should be further developed. The latter can also be seen from the writing, in some of the references, in (3b), "*f*" as a scalar (even if dependent on a vector), which is obviously wrong because the left hand-side of the equation is a vector.

A question arises regarding (3) – would it be possible for the function $R(.,.)$ to include $di/dt$, which does not contradict (3)? This question interests us in view of [31-33] where the $v(t)$-$i(t)$ dependence for the extremely significant fluorescent lamp appears as (see equation (8) below)

$$v = f(i, di/dt)i. \qquad (4)$$

This relation cannot be written as any $v(i)$, and the (weak) inductive feature of the lamp is obvious from the dependence of $v$ on $di/dt$. The classical monograph [22] gives the form (4), but in [22] '*f*' is claimed to be unknown. Introducing model $v = f(i) + g(\frac{di}{dt})$ with a singular $f(.)$, [41] makes it known.

**3.2  *Some hysteresis characteristics.*** Figures 2 and 3 show two hysteresis characteristics of real fluorescent lamps. The transfer from the high $dv/di$ to a low one, observed in these figures, is memristive behavior. We model such characteristics below, using the singular function sign[$i$]. Since for the lamp-circuit that converts the electrical energy into light, the averaged consumed power $P = <p(t)> = <v(t)i(t)>$ is the main parameter, it is important for a model to give correct values of $v$ at the *high* values of $i$, when $p(t)$ is relatively large. (Consider $P \sim \int p(t)dt$ .) Thus, though there is a significant deviation of sign[$i$] from the real $v(i)$ at *low* currents, this function is generally appropriate.

If, however, in a memristor application, one uses the lamp (or any other device with a hysteresis characteristic) as a voltage source for a load, then one has to remember that *per se* it is a very non-ideal voltage source, and that the voltage will be more stable when the current is not too small. Thus, such a memristor can require use of amplifiers to support modeling.

According to [1,2] a typical memristive hysteresis should disappear at high frequencies. This requirement does not seem to be necessary for applications (and in Section VIII we consider a hysteresis system not necessarily showing this disappearance), but this is, indeed, the situation [32] with the fluorescent lamps.

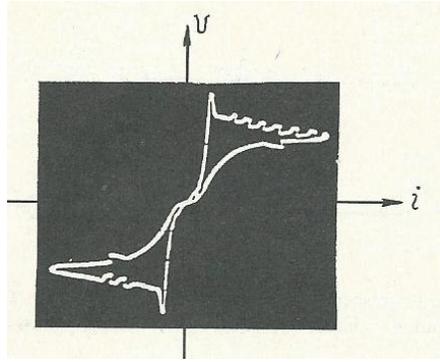

Fig 2: The (mainly) resistive hysteresis loop of a fluorescent lamp. The direction of the loop-movement is clockwise; the voltage obtains its highest value *before* the current, which is an inductive feature. The transfer from the high value of the differential resistance d*v*/d*i*, to the low value and backwards relates to the memristive behavior, but, as explained in the main text, use of the hysteresis features under load can require amplifiers, i.e. sources of energy (batteries).

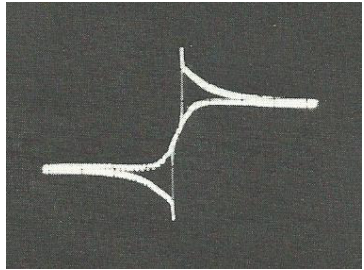

Fig 3: A loop for a fluorescent lamp with a smaller tube diameter, i.e. a smaller gas mass. In this case, the range of the values of d*v*/d*i* is very wide, but the voltage can be changed when a load is connected to the lamp in parallel fashion. That is, it is not easy to use the attributes of the memristive behavior.

The lamp's *inductive* feature is associated with the *clockwise* direction of its hysteresis *v-i* loop. (Indeed, *for a pure inductor*, having the Lissajous figure in the *i-v* plane as a circle, the voltage leads the current, i.e. the point ($i = 0$; $v = v_{max}$) comes in time before the point ($i = i_{max}$; $v = 0$).) The opposite direction of such a loop means capacitive behavior. See [32,20,21] for more details.

One observes that clearly any hysteresis characteristic relates to memory, because movement along the hysteresis loop is associated with *remembering* the previous position on the loop.

Another remarkable fact, stressed in [32,33], is that the energy associated with the lamp's inductance $L'$,

$$W_{L'} = L'\frac{i^2}{2}, \qquad (5)$$

is of an *electrostatic origin*. This energy is associated with the separation of positive and negative charges because of the bipolar diffusion. There is no significant magnetic flux associated with the lamp, justifying the rather large value of $W_{L'}$. *Thus, introducing di/dt into (4) we mathematically obtain an inductor, but its energy appears to be of a capacitor's nature. In diffusion processes, the current can be delayed with respect to the voltage, which is an inductive* *feature per se, agreeing with v ~ di/dt, or i ~ ∫v dt , but not having any relation to any significant magnetic flux and magnetic energy.* See also [34,35] suggesting the possibility of inductive features of the human-body impedance, because of possible diffusive processes in the body.

***3.3 The use of the singular time functions.*** Regarding the purely resistive $v(i)$-model of first approximation of the lamp, suggested by Figs 2 and 3,

$$v(i) = A\, sign[i] , \qquad (6)$$

with a constant $A$, not yet taking into account the hysteresis, it is natural to transfer to the relation between the *time functions*:

$$v(t) = A\, sign[i(t)], \qquad (7)$$

allowing the periodicity of $i(t)$ to be well used in the circuit's Fourier analysis. Indeed, for $i(t)$ without pauses where it would be identically zero, the $sign[i(t)]$ is a simple rectangular wave [31,32] that could also be written as $v(t) = A\, sign[\sin \omega t]$, with the period $T = 2\pi/\omega$ the same as that of the non-sinusoidal $i(t)$ satisfying in such a circuit the equation $i(t+T/2) = -i(t)$ and having only 2 zerocrossings per period. The use of (7) will be our focus.

Since, as is explained in [36], for many switched circuits and systems no characteristic exists, it is sometimes desirable to use from the very beginning models involving time-functions $v(t)$ and $i(t)$. Usage of such models is especially important when an element cannot be properly defined when taken separately, but only when it is inside a circuit. (Notice that the fluorescent lamp is such an element.) In fact, (3) is also similar to this, and see [37] for such considerations. In the following section we demonstrate how to transfer from (7) to a hysteresis characteristic, namely to

$$v(t) = A_1 sign[i(t)] + L'\frac{di}{dt} \qquad (8)$$

where the inductive term is added *and the factor before* sign[$i(t)$] *is increased*. This is a much more flexible description of the lamp than (7).

The range of d$v$/d$i$ in (6-8) is extremely wide, and we have to somewhat smooth the jumps of this singular function in order to be closer to the real situation, and to have time for using the high value of the derivative in an application. However, when speaking about power consumption, which is the inherent feature of any resistive element, we do find that (6) and (8) are good models. Observe also that, using (8),

$$P = \langle i(t)v(t)\rangle = \left\langle i(t)(A_1 sign[i(t)] + L'\frac{di}{dt})\right\rangle = A_1 \langle |i(t)|\rangle ,$$

because $i(t)$ and d$i$/d$t$ (as well as sign[$i(t)$] and d$i$/d$t$) are mutually orthogonal over large time intervals and in particular for periodic processes when the averaging can be done over one period. The relation $P \sim |i|$ is important for revealing the unusual scaling features of the lamp's power. See also [32,33] for analysis of the power.

IV. AN EXAMPLE OF ANALYZING HYSTERESIS CHARACTERISTICS

As shown below, the more detailed form of (8) is

$$v(t) = A(1+2\frac{L'}{L}) sign[i(t)] + L'\frac{di}{dt} \tag{9}$$

i.e. $A_1 = A(1+2\frac{L'}{L})$ where $L$ is the inductance of the ballast $B$ shown in Fig. 4, and $A$ keeps its sense of $v$ taken at $i_{max}$. (As is explained in [32] and referenced there, $A$ has a very stable value associated with the quantum features of the low-pressure gas discharge.) Thus, (6) can be seen as a limiting case of a symmetric hysteresis loop when the hysteresis parameter $L' \to 0$.

Equation (9) is relevant to the values of d$v$/d$i$ that can be obtained from the hystersis loop, and also to the element's power consumption. This equation is supported by the following observation. Both of the inequalities $L' > 0$ and $A_1 > A$ reflect the same hysteresis phenomenon, i.e. it should be $A_1 - A \sim L'$. Because of the dimensional (physical units) reasons $(A_1 - A)/A = kL'/L$ for some non-dimensional constant $k$, while $L$ must be a property of the ballast $B$, since there is no other inductance to serve as reference. Thus, $A_1 = A(1+ kL'/L)$.

The presence of the external inductance $L$ in the $v(t)$-$i(t)$ relation (9) supports the aforementioned opinion of [37] that it may be better to define an element not independently, but rather in the circuit context.

A systematic nonlinear theory of fluorescent lamp circuits based on the simplest model (6,7) is given in [31]. The original derivation of (9), given in [33], uses a separation of $i(t)$ into two parts different in their analytical properties (smoothness). This specific analytical method is very helpful in employing the singular models for such systems.

*4.1 The smooth and rough parts of the current function.*    Considering the circuit of Fig. 4, where '$e$' is the lamp, we write

$$\begin{aligned} i(t) &= \hat{L}_1[v_{in}(t)] - \hat{L}_2[v(t)] \\ &= \hat{L}_1[v_{in}(t)] - A\hat{L}_2[sign[i(t)]] \end{aligned} \tag{10}$$

where $\hat{L}_1$ and $\hat{L}_2$ are *linear integral operators* representing the steady state output-current response of $B$ to its inputs. (When considering $B$ *per se*, we can, according to the basic substitution theorem [26,27], interpret '$e$' as an ideal voltage source, and $v$ as an independent input of the 2-port.) Although the last term of (10) shows that we start with model (6,7) for the lamp, -- as the essence of the iterative step, (10) will be immediately used for correcting this model.

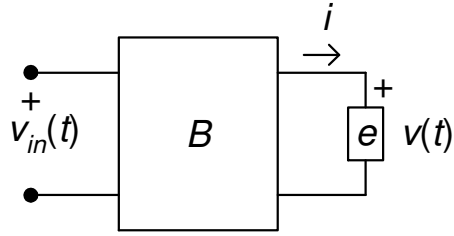

Fig.4: The lamp 'e' is connected via linear "ballast" $B$ including a large inductance $L$ providing uninterrupted current $i(t)$. The ballast can be a series circuit, becoming a 1-port, and then for the integral operators in (10) $\hat{L}_1 = \hat{L}_2$.

The next step is the aforementioned separation of the lamp's current into two parts, $i(t) = i_1(t) + i_2(t)$, so that $i_2(t)$ is more singular, i.e. its Fourier coefficients are slowly decreasing with their number, namely $O(1/n^2)$ (consider the second term in (10)), while Fourier coefficients of the smoother $i_1(t)$ are $O(1/n^m)$, $m > 2$. The separation of the $i(t)$ into two functions different in the orders of the Fourier coefficients, i.e. by the functions' degrees of singularity, appears to be very straightforward and effective [33] in the nonlinear theory of the lamp circuits. Using the degree of freedom provided by the separation of $i(t)$ into two parts, we associate two lamp's inductances, $L_1'$ and $L_2'$, with these currents, as:

$$v(t) = A\,sign[i(t)] + L_1' \frac{di_1}{dt} - L_2' \frac{di_2}{dt} . \tag{11}$$

Rewriting (10) as

$$\frac{di}{dt} = \frac{d}{dt}\hat{L}_1 v_{in}(t) - A\left(\frac{d}{dt}\hat{L}_2 sign[i(t)] - \alpha\,sign[i(t)]\right) - \alpha A\,sign[i(t)] \tag{10a}$$

using a constant $\alpha$, we define $i_1(t)$ and $i_2(t)$ by discovering $\alpha$ from the requirement that the terms with Fourier coefficients of order $O(1/n)$ (which in $i(t)$ becomes $O(1/n^2)$) are cancelled in the parenthetic term, i.e. the parenthetic term becomes a smooth (continuous) function related, together with $\frac{d}{dt}\hat{L}_1 v_{in}(t)$, to $di_1/dt$. Thus, for such $\alpha$, we conclude from (10a) that for the more singular $i_2$

$$\frac{di_2}{dt} = -\alpha A\,sign[i(t)] . \tag{12}$$

Since according to eq' (13) below, the Fourier coefficients of $\frac{di_2}{dt} \sim sign[i(t)]$ are $O(1/n)$, those of $i_2(t)$ are indeed $O(1/n^2)$. It appears [33] that $\alpha$ is the converse value of the *asymptotic inductance L* of $B$, that is

$$\alpha = \frac{1}{L} \equiv \lim_{n\to\infty} n\omega \,|Y(n\omega)|,$$

where $Y(.)$ is the *admittance function* [26,27] of $B$.

Substituting into (11) first

$$\frac{di_1}{dt} = \frac{di}{dt} - \frac{di_2}{dt}$$

and then using (12) and that $\alpha = 1/L$, we finally obtain (11) as

$$v(t) = A(1 + \frac{L_1' + L_2'}{L})\,sign[i(t)] + L_1' \frac{di}{dt}.$$

Setting for simplicity $L_1' = L_2'$, and denoting this common value as $L'$, we obtain (9).

The intermediate use, of the two lamp's inductances, $L_1'$ and $L_2'$, suggests completing the series (voltage-hardlimiter and inductance) model of the lamp in [32,33] through use of a model with a perfect transformer characterized by two inductive parameters. This completion has not yet been developed.

The hysteresis model (9) used in [32,33], explains the main experimentally observed features of the real fluorescent lamp circuits clearly, and the wide interest in [4-19] to hysteresis characteristics makes the results of [32,33] relevant to the theory of memristive systems.

The circuit-theory essence of the singular modeling is that when passing on from (6) to (7) and then to (9), we immediately pass on from a characteristic $v(i)$, to a connection $v(t)$-$i(t)$, i.e. not using a map $i \to v$ of a number on a number, but a system-operator map $i(t) \to v(t)$ of a function on a function.

The jumps of the singular term provide the jumps of the value of the differential resistance, which is an idealization of the switch operation of [4]. The instants where these jumps occur and the information that these instants carry are associated with some specific system nonlinearity and with a useful interpretation of the state equations of a switched systems.

V  THE "ZEROCROSSING NONLINEARITY" AVOIDING THE V(I) CHARACTERISTICS

Zerocrossings, or levelcrossings, $\{t_k\}$ of a time function of a system under study, are the basic analytical parameters connecting the use of the singular time-functions with the very important concept of *zero- (level-) crossing nonlinearity*.  Even if only in view of model (3), the nonlinearity of a resistive system should not always be associated only with the curviness of some $f(i)$. New ways of presentation of a system's nonlinearity have to be found, and studies [36,39] show such a way for switched (singular) systems, using $\{t_k\}$, which can be useful for the memristive systems' theory.

The previously mentioned nonlinear theory of fluorescent lamp circuits can be formulated using the explicit introduction of the zerocrossings into the system's mathematical description. In the case when $i(t)$ is an uninterrupted (i.e. not having intervals of nonzero length where it would be identically zero) and a zerocrossing periodic function, according to (7), $v(t) \sim \text{sign}[i(t)]$ is a square wave time function provided by the Fourier series (see, e.g., [38]):

$$\text{sign}[i(t)] = \frac{4}{\pi} \sum_{1,3,5,\ldots}^{\infty} \frac{\sin[n\omega(t-t_1)]}{n}; \quad t_1 : i(t_1) = 0 \qquad (13)$$

where $t_1$ (modulo $T$, $T = 2\pi/\omega$) is a -/+ zerocrossing of $i(t)$. When considering more complicated waveforms, the whole set $\{t_k\}$ of the zerocrossings belongs to one period has to be used, with any even number of the zerocrossings.

The use of the zerocrossings allows one, via system analysis [31-33,39], to present the current as a known function of time and its own zerocrossings, $i = F(t, \{t_k\})$, and then the symbolic equation $i(t_k) = 0$ becomes the constructive equation $F(t_p, \{t_k\}) = 0$, $p = 1, 2, \ldots$, from which the zerocrossings, and thus also $i(t)$ can be found. It is important to see that, generally, for a nonlinear system, $\{t_k\}$ *cannot be prescribed*, since the function to which $\{t_k\}$ relates may be a state variable to be found; that is, the zerocrossings are introduced as some initially unknown parameters. This can be seen in (13), where $t_1$ is still unknown, but introduction of this parameter is the essence of working with the singular characteristic (6,7), because *only it* is unknown [41,48].

*5.1 The logical aspect*. Let us compare the zerocrossings to any other analytical parameters that may appear in the equations that describe a system, e.g. to the coefficients of a differential equation. For the coefficients, we have three cases:

a. *these coefficients are constant*

b. *they are known time functions*

c. *they depend on the state variables to be found*. In this case, they are, in fact, not "coefficients", but it is *useful* to see them as some system's structural parameters depending on the state variables.

In cases "a" and "b" of a linear (respectively time invariant and variant, LTI and LTV) system, the coefficients are prescribed a priori, and in case "c" (of an NL system) they cannot be prescribed.

This is also the logical situation with $\{t_k\}$. When $\{t_k\}$ can be prescribed, the system is linear, either LTI (when $\{t_k\}$ are constant under any test of linearity), or LTV. When $\{t_k\}$ cannot be prescribed, we have the "zerocrossing nonlinearity". (Observe that while zerocrossings preserve their precise definitional sense also when they depend on unknown functions, it is, strictly speaking, not the situation with the coefficients of an equation; thus, when introduction of the "zerocrossing nonlinearity" is possible, it can have some advantage.)

With these observations, we return to Fig. 4 and take

$$v_{in}(t) = U\xi(t)$$

where $U$ is a scaling parameter, and $\xi(t)$ is a $T$-periodic wave defining the waveform of the input function and the position of the time origin, i.e. the point $t = 0$. Changing $U$, we can observe nonlinear effects in this driven circuit. For instance, if it is observed (as in [32,33,42]) that the circuit's *steady-state* power consumption is *not* proportional to $U^2$, then the circuit is nonlinear. Indeed, only for a linear system the losses have the form $Ri_R^2 \sim U^2$.

The nonlinearity of the circuit can take two very different forms. As the usual case, it may be that the zerocrossings of $i(t)$ are dependent on $U$, i.e. are moved with respect to a zero of $\xi(t)$, when $U$ is changed. It also may be that the zercrossings are independent of $U$, i.e. unmoving, constant, which can be provided, in some range of $U$, by a special synthesis of $B$ [40-42].

***5.2 The scaling aspect of nonlinearity and the case of affine nonlinearity.*** In order to illustrate the classification aspect, consider a simple version of the fluorescent lamp circuit described by the equation

$$L\frac{di}{dt} + A \, sign[i(t)] + \frac{1}{C}\int^t i(\lambda)d\lambda = v_{in}(t), \qquad (14)$$

i.e.

$$\left(L\frac{di}{dt} + \frac{1}{C}\int^t i(\lambda)d\lambda\right) + \left(\frac{4A}{\pi}\sum_{1,3,5,\ldots}^{\infty}\frac{\sin[n\omega(t-t_1)]}{n}\right) \qquad (15)$$
$$= U\xi(t).$$

If $t_1$ depends on $U$, then the second parenthetic term in (15) is obviously nonlinear (the zerocrossing nonlinearity). If $t_1$ is independent of $U$, then this term presents a known time function. However, in this case the system is *also* nonlinear, in the sense of the *affine nonlinearity*. Indeed, denoting the term with the sum as $-f(t)$, we can rewrite (15) as

$$(\hat{L}i)(t) = U\xi(t) + f(t), \qquad (15a)$$

with a linear operator $\hat{L}$, and clearly $U \to kU$, with a constant $k$, does *not* lead to $i \to ki$.

Thus, in accordance with the fact that the fluorescent lamp included in the circuit is a nonlinear element, it appears that the map $U \to i$ (or $v_{in} \to i$) in (14) is always nonlinear. However, the appearance of the case of affine nonlinearity is an interesting specificity of the zerocrossing nonlinearity, and it would be interesting to reveal such different cases of the memristive systems.

On the basis of circuit theory [26,27], the term $f(t)$ in (15a) is an application of the basic substitution theorem, in which a passive element is replaced by an ideal source, but such an application is very rarely as constructive as it is here.

## VI. THE ZEROCROSSING NONLINEARITY AND THE STATE-EQUATIONS

In order to formalize the use of the parameters $\{t_k\}$ and better show the physical sense of this use, let us write, following [39,36,35], the state equations in the "structural form", i.e. not as

$$\frac{d\vec{x}}{dt} = \vec{F}(t,\vec{x},\vec{u}), \qquad (16)$$

but as

$$\frac{d\vec{x}}{dt} = [A(t,\vec{x})]\vec{x} + [B]\vec{u} \qquad (17)$$

where the structure is given by matrices [A] and [B] (focusing only on [A]). For [A] and [B] independent of $\vec{x}$, the system is clearly linear, and one sees from (17) that the definition of nonlinearity of a system via the dependence of the system's structure on $\vec{x}$ (or on the inputs [39]) is a good pedagogical and heuristic definition.

Indeed, the heuristically important concept of structure is seen in (17) and not in (16). Thus, for instance, the nonlinearity of hydrodynamic equations (when they are still unwritten) is directly seen based on the fact that the velocity field of a liquid flow is *both* the variable to be found and the structure of the flow (system), which is the philosophy of (17) and not of (16). That is, the liquid flow is an "[A(x)]-system". (See [39] for additional details and examples.)

In order to move from the very general form (17) to the zerocrossing nonlinearity relevant to the switched systems, we need to observe that for *any* switched system, LTV or NL, we have

$$\frac{d\vec{x}}{dt} = [A(t,\{t_k\})]\vec{x} + [B]\vec{u}, \qquad (18)$$

where $\{t_k\}$ are the instants of the switch occurring in the system. If $\{t_k\} = \{t_k(\vec{x})\}$, i.e. the switching instants are defined by at least by one of the functions (state variables) to be found, then $[A(t,t_k(\vec{x}))]$ is some $[A(t,\vec{x})]$, i.e. (18) is the NL case (17).

The zerocrossing nonlinearity may be quite naturally included in (3); one can consider, for instance, the form

$$v = R(\vec{x},i)i$$
$$\frac{d\vec{x}}{dt} = \vec{f}([E(t,\{t_k(\vec{x})\})],i)$$

with some matrix [E], or the simpler form

$$v = R(\vec{x},i)i$$
$$\frac{d\vec{x}}{dt} = [A(t,\{t_k\})]\vec{x} + [B]\vec{u}$$

where $\vec{u}$ can be $v$ or $i$.

The following lucid example of the zero-crossing nonlinearity strongly supports this view on the state equations and the definition of nonlinearity, which is a line of thought relevant to the memristive systems.

VII. TWO LINEAR (SUB)SYSTEMS CONTROLLED BY A NONLINEAR SWITCHING [40]

Reference [40] discusses two topologically similar linear (sub)circuits having only one different element. Switching from one (sub)circuit to another means that the (similarly defined) outputs of the circuits are used in turn.

This use of two circuits is the same as the use of *one circuit* having (only) one of its elements switched, from time to time, from one of the two given values to the other and back. See Fig.5. That is, instead of speaking about different linear (sub)systems, we can speak, as for any usual switched system, about different states of the same system, which is a more compact structure. (In very simplified form, one can imagine a system including a switch, which is open in one state, and closed in the other state; then the transfer $R_1 \leftrightarrow R_2$ in Fig. 5, becomes $0 \leftrightarrow \infty$.)

Thus, we can consider the system of [49] in view of (18), and the question is how [40] defines the switch, i.e. whether we have $\{t_k\}$ prescribed, or they depend on the state variables, $\{t_k\} = \{t_k(\vec{x})\}$. In the latter case, we can transfer here from (18) to (17) having an NL system of the $[A(t,\vec{x})]$-type.

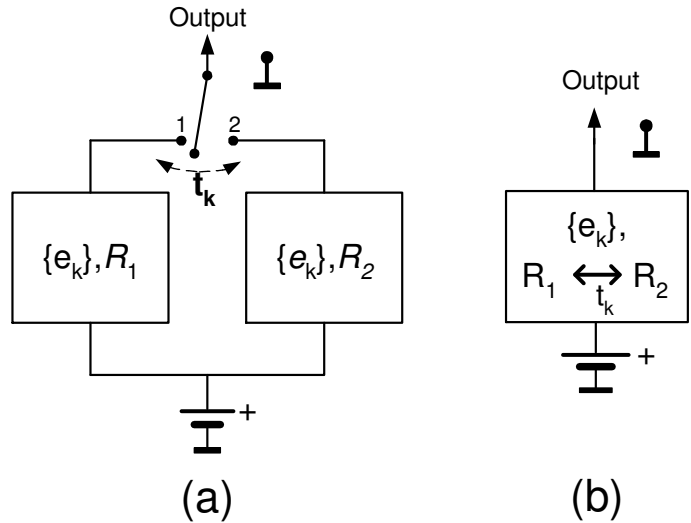

Fig.5: (**a**) The version of [40] with two linear subsystems; (**b**) The equivalent version in terms of (18). The set $\{e_k\}$ denotes all the circuit elements other than $R_1$ and $R_2$. Since the actual operation of the circuit, i.e. the transfer from one subcircuit to the other, is such that $\{t_k\} = \{t_k(\mathbf{x})\}$, (18) is (17), the whole system is NL, and the resulting chaos in [40] is not surprising.

The switching is done in [40] when a *state variable* crosses some given level, which is checked online, using a computer. This means that $\{t_k\} = \{t_k(\vec{x})\}$, and the system is NL. Classification of the methods of obtaining $\{t_k(\vec{x})\}$ in electronics would be very useful for nonlinear circuit theory.

The fact that using the zerocrossing nonlinearity $\{t_k\} = \{t_k(\vec{x})\}$ one can obtain any nonlinear effect is also supported by work [41] that demonstrates a chaotic process obtained by a "mirror reflection from time axes" of a function $x(t)$ at its zerocrossings $\{t_k\}$. That is, the equations $dx/dt(t_k^-) = - dx/dt(t_k^+)$ are forced, making $t_k$ be dependent on $x$, while the circuit is otherwise defined using an LTI equation.

We continue with the use of hysteresis characteristics, now in a very different circuit context.

## VIII.  HYSTERESIS CHARACTERISTICS USING POWER-LAW CIRCUITS

Among the resistive circuits, the *power-law* circuits [43-46] composed of similar elements, all with the same *power-law* characteristic

$$v = D_\alpha i^\alpha \qquad (19)$$

where the constant $D_\alpha$ has proper physical units, are of interest here. We shall write (19) using two constants, $v_o$ and $i_o$, having the usual units of volts and amperes:

$$\frac{v}{v_o} = \left(\frac{i}{i_o}\right)^\alpha . \qquad (20)$$

Obviously, the point $(i_o, v_o)$ belongs to the $v(i)$-curve for any $\alpha$. (Consider also Fig. 6 below.)

At the limiting case of $\alpha \to \infty$, (20) becomes a *current hardlimiter* at the value $i \equiv i_o$:

$$v \underset{\alpha \to \infty}{\to} \begin{cases} 0, & 0 \leq i < i_o \\ \infty, & i_o < i \end{cases}.$$

Since, furthermore, from (20)

$$\frac{i}{i_o} = \left(\frac{v}{v_o}\right)^{1/\alpha},$$

we have

$$i \underset{\alpha \to 0}{\to} \begin{cases} 0, & 0 \leq v < v_o \\ \infty, & v_o < v \end{cases},$$

that is, for $\alpha \to 0$, we obtain the opposite case of the *voltage hardlimiter* at the level of $v \equiv v_o$.

Thus, for $\alpha \gg 1$, we deal with a characteristic close to a *current hardlimiter*, and for $\alpha \ll 1$ with a characteristic close to a *voltage hardlimiter*.

Using such (electronically realized) characteristics as

$$v = \begin{cases} D_{\alpha_1} |i|^{\alpha_1} \, sign[i], & \dfrac{di}{dt} > 0 \\ D_{\alpha_2} |i|^{\alpha_2} \, sign[i], & \dfrac{di}{dt} < 0 \end{cases} \qquad (21)$$

where, generally, $\alpha_2 \neq \alpha_1$, one can create memristive 1-ports, with hysteresis characteristics which for $\alpha_{1(2)} \gg 1$, or $\alpha_{1(2)} \ll 1$, will be close to the hardlimiter characteristics. The reason why this closeness, i.e. use of the extreme values of $\alpha_{1(2)}$, is important, is associated with the switching applications of [4]. Figure 6 illustrates (21), allowing one to consider $dv/di$. For $\alpha_2 > \alpha_1$ we have an inductive loop, and for $\alpha_1 > \alpha_2$, -- a capacitive one.

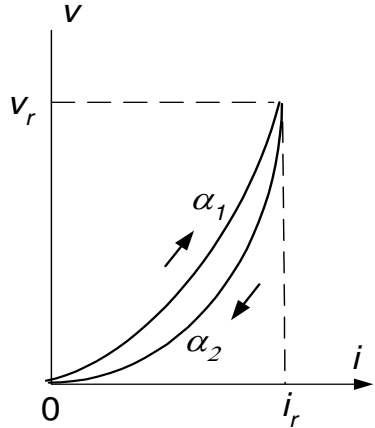

Fig. 6: An illustration to (21), for $\alpha_2 > \alpha_1$ (an inductive loop), presented in the interval $0 < i < i_r$ of the whole interval $-\infty < i < \infty$ where $v(i) = -v(-i)$. The larger $\alpha_2/\alpha_1$ is, the wider the range of $dv/di$ is, and the better the memristor is, in the sense of [4].

The return point $(i_r, v_r)$ (which is $(i_o, v_o)$ in the notations of (20)) of this rough "eye-type" loop is found from the equation (we take $i > 0$)

$$D_{\alpha_1} i_r^{\alpha_1} = D_{\alpha_2} i_r^{\alpha_2},$$

i.e.

$$i_r = \left(D_{\alpha_2}/D_{\alpha_1}\right)^{\frac{1}{\alpha_1-\alpha_2}}, \qquad (22)$$

and $v_r$ is then easily found from (21).

If $\alpha_2 = \alpha_1 + 1$, then from (22) $i_r = D_{\alpha_1}/D_{\alpha_2}$.

The case of $v \sim i^{1-\varepsilon \cdot \text{sign}[di/dt]}$, $\varepsilon > 0$, (i.e. $\alpha_2 = \alpha_1 + 2\varepsilon$) gives a hysteresis loop that for $\varepsilon \to 0$ becomes a linear resistor characteristic.

Contrary to the case of a fluorescent lamp, this hysteresis need not disappear at high frequencies, which should be good circumstance for applications.

## IX. CONCLUSIONS AND FINAL REMARKS

Memristive circuits are very interesting for physical and circuit analyses. The memristor with its *RLC* features is, first of all, a logical quintessence of the physical reality of any resistor. Since any use of the variables *v* and *i* introduces a capacitance and an inductance, *any resistor* described using some *v*(*i*) characteristic is associated with the energy storing elements. This physical aspect, related to the very foundation of circuit theory, has to be always kept in mind, and it also has to be faced that the invention of the memristor unexpectedly presented resistor as the theoretically most complicated circuit element.

Motivated by works [2,11,25,32] with their intermediate discussion of some mainly-resistive nonlinear elements, interpreted as memristors, with hysteresis characteristics whose (usually weak) *C* and *L* features are obvious, we have had to survey some nontrivial analytical details of a nonlinear theory of fluorescent lamp circuits, which should be useful for a wider analysis of elements with hysteresis, relevant to the memristors. These details show how to avoid the use of the unique *v*(*i*) characteristics as quickly as possible.

In particular, we show that a connection between the voltage and the current functions of the form

$$v(t) = f(i(t)) + g(di/dt)$$

where *f*(.) is singular, can be also suitable for a modeling. The singularity introduces zero- (level) crossings of time functions, and the nonlinearity of a circuit is well expressed when the zero- (level) crossings depend on at least one state-variable. The singularity is naturally associated with switched systems, when the points of the singularity become switching instants. The information included in the dependences $\{t_k = t_k(\vec{x})\}$ can be compared with the information given by $\vec{x}$ in (3). One sees that the topic of the zero- (level-) crossing nonlinearity essentially enriches the basic notion of nonlinearity, supporting the position that circuit nonlinearity need not be characterized by a curviness of the characteristic of an included element. This is a constructive point of view on nonlinearity, which should be taught in basic EE education.

Regarding the fluorescent lamps' circuits *per se*, it has to be commented that for any *linear* model of the lamp, it is unclear why the ballast is at all needed, and the actually problematic high harmonic currents can not be estimated. Unfortunately, the theory of [31-33], never included in any textbook, is generally unknown to power-system engineers, and the absolutely unsatisfactorily linear *R-L* model of the lamp is still used. The high harmonic currents are not calculated, they are measured by the electrical company, and then one pays a fine if these currents are too high. Hopefully, the analysis of the memristive systems will be useful for the gas discharge devices (that have recently found application also in the field on plasma antennas [47-54]) and for heavy-current problems in their wide scale.

Finally, the unusual power-law circuits of Section VIII allow one to create specific hysteresis characteristics of the eye (crescent, lune) shape. This should require some switching of the elements, but such circuits should not have the disadvantage of being strongly frequency dependent in their physical features as is the case with such discharge (diffusion) devices as florescent lamps and neon bulbs.

Works [1-22,25] are recommended for completion of the physical side. We hope that the present heuristic arguments can contribute to the *logical position* of the memristor in circuit theory, to the analysis of some associated electromagnetic field problems, to seeking new applications, and to the pedagogical aspect.

ACKNOWLEDGMENT

I am grateful to Jean-Marc Ginoux for kind encouragement and some good old references, e.g. [55], to Chandler Davis for some style corrections.

REFERENCES

[1] L.O. Chua, "Memristor -- The Missing Circuit Element", IEEE Transactions on Circuit Theory, VOL. CT-18, NO. 5, SEPTEMBER 1971, pp. 507-519.

[2]  L.O. Chua, & Kang S. M. [1976] "Memristive devices and systems," *Proceedings of the IEEE* **64**, 209–223.

[3]  L.O. Chua, "Device Modeling Via Basic Nonlinear Circuit Elements", IEEE Transactions on Circuit Theory, 1980, vol. CAS-27, no. 11, p. 1014-1044.

[4]  D.B. Strukov, G. S. Snider, D.R. Stewart, and R.S. Williams, "The missing memristor found," *Nature* 453, 80–83 (2008).

[5]  D. Jeltsema, A. Do`ria-Cerezo, "Port-Hamiltonian Formulation of Systems With Memory", Proceedings of the IEEE, 2011, vol. 100, no. 6, p. 1928-1937.

[6]  P. Meuffels, R. Soni, (2012). "Fundamental Issues and Problems in the Realization of Memristors". arXiv:1207.7319 [cond-mat.mes-hall].

[7]  I. Valov, E. Linn, S. Tappertzhofen, S. Schmelzer, J. van den Hurk, F. Lentz, & R. Waser, "Nanobatteries in redox-based resistive switches require extension of memristor theory", *Nature Communications* **4** (Article number: 1771, doi:10.1038/ncomms2784) 23 April 2013.

[8]  M. Di Ventra; Y. V. Pershin, (2013), "On the physical properties of memristive, memcapacitive and meminductive systems", *Nanotechnology* **24** (25): 255201,arXiv:1302.7063, Bibcode:2013Nanot..24y5201D, doi:10.1088/0957-4484/24/25/255201, PMID 23708238

[9]  L. B. Kish, C. G. Granqvist, (2014). "DEMONS: MAXWELL'S DEMON, SZILARD'S ENGINE AND LANDAUER'S ERASURE–DISSIPATION".arXiv:1412.2166 [physics.gen-ph].

[10] V. A. Slipko; Y. V. Pershin, M. Di Ventra, (2013), "Changing the state of a memristive system with white noise", Physical Review E **87**: 042103

[11]  L. O. Chua, "If it's pinched it's a memristor," Semicond. Sci. Technol., vol. 29, pp. 1-42, 2014.

[12]  N.; Hashem, S. Das, (2012), "Switching-time analysis of binary-oxide memristors via a non-linear model", *Applied Physics Letters* **100** (26): 262106, Bibcode:2012ApPhL.100z2106H, doi:10.1063/1.4726421, retrieved 2012-08-09.

[13]  E. Linn; A. Siemon; R . Waser,; S. Menzel, (23 March 2014). "Applicability of Well-Established Memristive Models for Simulations of Resistive Switching Devices". *Circuits and Systems I: Regular Papers, IEEE Transactions on* **61** (8): 2402–2410.

[14]   L. Czerwiński, M.J. Ogorzalek: "Study of Bifurcation Phenomena in Memristor Oscillators", Proc. NOLTA 2011, Kobe, Japan, September 2011, pp.4-7.

[15] Z. Galias, "Computer assisted proof of chaos in the Muthuswamy-Chua memristive circuit", NOLTA, IEICE, July 1, 2014, pp. 309-319.

[16] S. Kvatinsky, E. G. Friedman, A. Kolodny, and U. C. Weiser, "The Desired Memristor for Circuit Designers", Second quarter 2013, Circuits and Systems Magazine, pp. 17-22.

[17] Y. Ho, G. M. Huang, P. Li, "Nonvolatile Memristor Memory: Device Characteristics and Design Implications", *ICCAD'09,* November 2–5, 2009, San Jose, California, USA, pp. 485-490.

[18] Ch. K. Volos, I. M. Kyprianidis, I. N. Stouboulos, E. Tlelo-Cuautle, and S. Vaidyanathan, "Memristor: A New Concept in Synchronization of Coupled Neuromorphic Circuits", Journal of Engineering Science and Technology Review 8 (2) (2015) 157 – 173. (Special Issue on Synchronization and Control of Chaos: Theory, Methods and Applications.)

[19]  E. D. Torre, "*Magnetic Hysteresis*", Wiley-IEEE Press, New York, 2000.

[20] M. E. Lines and A. M. Glass, '*Principles and Applications of Ferroelectric and Related Materials*", Oxford, U.K.: Clarendon, 1977.

[21] J. C. Burfoot and J. W. Taylor, "*Polar Dielectrics and Their Applications*", New York: Macmillan, 1979.

[22] J. F. Waymouth, "*Electrical Discharge Lamps*" Cambridge: The MIT Press, 1978.

[23]  F. Klein, "Forlesung uber die Entwicklung der Mathematik im 19. Jahrhundert", Springer, Berlin, 1926.

[24] E. Gluskin, "The electromagnetic 'memory' of a dc-conducting resistor: a relativity argument and the electrical circuits", Manuscript, 2010, arXiv:1005.0997

[25]  J.-M. Ginoux and B. Rossetto, "The singing arc: the oldest memristor? In "Chaos, CNN, Memristors and Beyond", Andrew Adamatzky et  Guanrong Chen (Ed.) (2013), pp. 494-507.

[26]  Ch.A. Desoer. E.S, Kuh, "*Basic Circuit Theory*", McGraw Hill, New York, 1969.

[27]  L.O. Chua, Ch. A. Desoer, E.S. Kuh, "*Linear and nonlinear circuits*", McGraw-Hill, 1987.

[28]  M.N.O. Sadiku, "*Elements of Electromagnetics*", Oxford University Press, New York, London, 2001.

[29] P.M. Morse and H. Feshbach, "*Methods of Mathematical Physics*", McGraw-Hill, New York, 1953.

[30]   A.N. Kolmogorov, "Some fundamental problems in the approximate and exact representation of functions of one or several variables", Proc. III Math. Congress USSR vol. 2 MGU Press, Moscow, 1956, 28-29.  (See also in "Selected Works by Kolmogorov", Kluwer Acad. Publ., London, and there the comments by V.I. Arnold.)

*E. Gluskin* "**The physical and circ.-theor. significance of the Memristor : Full version**".arXiv:1602.02744 [cs.ET]    22[31] E. Gluskin, "The nonlinear theory of fluorescent lamp circuits", Int'l. J. of Electronics, 63, 1987 (687-705).

[32] E. Gluskin, "The fluorescent lamp circuit", Circuits & Systems Expositions, IEEE Transactions on Circuits and Systems, Part I:   Fundamental Theory and Applications, 46(5), 1999, pp. 529-544.

[33]  E. Gluskin, "The power consumed by a strongly nonlinear element with a hysteresis characteristic fed via a periodically driven LC circuit", J. of the Franklin Institute, 328(4), 1991 pp. 369-377.

[34]  E. Gluskin "On the human body's inductive features: A comment on 'Bioelectrical parameters…' by A.L. Lafargue et al. "Bioelectromagnetics, 24(4), 2003 (292-293).

[35]  E. Gluskin,"Some motivating examples for teaching electrical engineering students", Far East Journal of Mathematical Education, vol. 6, no.1, 2011, pp. 65-80.

[36]  E. Gluskin, "Switched systems and the conceptual basis of circuit theory", Open Column – IEEE Circuits and Systems Magazine, 3rd quarter 2009 (pp. 56, 58, and 60-62).

[37]   A. Borys, On In-Network and Other Types of Amplifier Descriptions for Nonlinear Distortion Analysis, Intl. Journal of Electronics and Telecommunications, vol. 56, no. 2, 2010, pp. 177-184.

[38]    H.B. Dwight, "*Tables of integrals and other mathematical data*", New York, Macmillan, 1961 (see item 416.01)

[39] E. Gluskin, "Structure, Nonlinearity, and System Theory" (2015), International Journal of Circuit Theory and Applications (CTA), vol. 43, issue 4, pp.  524-543. DOI: 10.1002/cta.1960.

[40]   X. Liu, K-L. Teo, H. Zhang and G. Chen, "Switching control of linear systems for generating chaos", Chaos, Solitons and Fractals, 30 (2006), 725-733.

[41]   H. Isomaki, J. von Boehn and R. Raty, "Devil's attractors and chaos of a driven impact oscillator", *Phys. Lett. A*, 1985, **107**A(8), 343-346.

[42] E. Gluskin, E. Tsirbas, I. Kateri, F.V. Topalis, "Use of logarithmic sensitivity in power system analysis: The example of lighting circuits (hot filament, LED and fluorescent lamp circuits)", IET Science, Measurement and Technology, vol. 7, no. 6, pp. 297–305, 2013.

[43]  E. Gluskin, "One-ports composed of power-law resistors", IEEE Trans. on Circuits and Systems II: Express Briefs, 51(9), 2004 (464-467).

[44]   E. Gluskin, "On the symmetry features of some electrical circuits", Int'l. J. of Circuit Theory and Applications – 34, 2006 (637-644).

[45] E. Gluskin, "An approximate analytical (structural) superposition in terms of two, or more, α- circuits of the same topology: Pt.1 – description of the superposition", Journal of Engineering and Journal of Engineering Science and Technology Review (JESTR), 6 (4) (2013) 33-44, (Special Issue on Recent Advances in Nonlinear Circuits: Theory and Applications.)

[46] E. Gluskin, "f-connection: a new circuit concept", 24th IEEE Convention of Electrical and Electronics Engineers in Israel – Electricity, 2006, Israel, (110-114)

[47] D. C. Jenn, "Plasma Antennas: Survey of Techniques and the Current State of the Art", Report NPS-CRC-03-001 (NAVAL POSTGRADUATE SCHOOL), September 29, 2003, San Diego, CA.

[48] J. Hettinger, "Aerial Conductor for Wireless Signaling and Other Purposes," Patent number 1,309,031, July 8, 1919.

[49] W. Manheimer, "Plasma Reflectors for Electronic Beam Steering in Radar Systems," *IEEE Transactions on Plasma Science*, vol. 19, no. 6, December 1993, p. 1228-1234.

[50] A. Moroney & M. Mehta, "Plasma antennas", International Journal of General Engineering and Technology (IJGET), ISSN(P): 2278-9928; ISSN(E): 2278-9936, Vol. 2, Issue 5, Nov 2013, 9-16.

[51] P. Darvish, "Plasma antenna by using fluorescent lamp and cylindrical couple", Iran Patent #79451 (05/2013)

[52] H. Ja'afar, M.T.B. , A.N.B. Dagang,. H.M. Zali, N.A. Halili, "A Reconfigurable Monopole Antenna With Fluorescent Tubes Using Plasma Windowing Concepts for 4.9-GHz Application", Plasma Science, IEEE Transact., Vol. 43 Issue 3, pp. 815 – 820.

[53] T. Anderson, "Multiple Tube Plasma Antenna, Patent number 5,963,169, October 5, 1999.

[54] E. Norris, T. Anderson, and I. Alexeff, "Reconfigurable Plasma Antenna," Patent number 6,369,763, April 9, 2002

[55] Ph. Le Corbeiller, "The Nonlinear Theory of the maintenance of oscillations," *Journal Inst. of Electrical Engineers*, 79, 361-378, 1936.

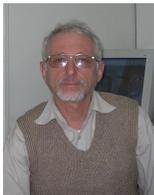

**Emanuel Gluskin** received the B.S. and M.S. degrees in physical engineering from Leningrad Polytechnic Institute (now Saint-Petersburg Technical University) in 1974, and the Ph.D. degree in electrical engineering from Ben Gurion University, Beer Sheva, in 1990.

He has industrial, research-authorities', and academic-teaching experiences. He published in physics, electrical engineering, and mathematical journals, introducing some new concepts, for instance, the most relevant here "zerocrossing nonlinearity", the "pause states" found in some oscillatory processes using symmetry argument (e.g. NOLTA, July 2014), a use of the inherent fractal nature of 1-ports, and a nonlinear transform, named "$\psi$-transform", for signal analysis.